\documentstyle[12pt]{article}
\begin{document}

\author{{\it Efrain J. Ferrer\thanks{%
E-mail : ferrer@fredonia.edu}\ and Vivian de la Incera\thanks{%
E-mail: incera@fredonia.edu}} \and SUNY at Fredonia, Fredonia, NY 14048, USA}
\title{{\bf REGGE TRAJECTORIES OF THE CHARGED STRING IN A MAGNETIC BACKGROUND}}
\date{July, 1995\\
{\bf SUNY-FRE-95-03}}
\maketitle

\begin{abstract}
The set of Casimir operators associated with the global symmetries of a
charged string in a constant magnetic background are found. It is shown that
the string rest energy can be expressed as a combination of these
invariants. Using this result, the Regge trajectories of the system are
derived. The first Regge trajectory is given by a family of infinitely many
parallel straight-lines, one for each spin projection along the magnetic
field.
\end{abstract}

One main problem of string theories is that up to now nobody knows how the
higher dimensions curl up and therefore what is the actual string vacuum. A
perturbative approach is insufficient to address this issue. On the other
hand, at present we lack of an adequate formulation of the theory that would
allow us to develop non-perturbative techniques to study it.

Given such an awkward situation, one ought to look for alternative ways to
obtain some understanding of the ground state properties. Exactly solvable
models are one interesting option \cite{hor-tse}. These are models of
strings in non-trivial background fields constructed under the requirement
of the conformal invariance of the theory. Although these models are
obtained by using an expansion in powers of the derivatives of the fields,
their solutions are exact in the sense that they contain all orders in $%
\alpha ^{\prime }$. In this way they provide us nonperturbative information
about the system and one expects they might be a good guide to discover the
hidden symmetries of the theory.

The simplest exactly solvable model is given by an open bosonic string in a
constant Abelian electromagnetic background. Since an open string interacts
with an electromagnetic background field through the charges attached at the
string ends, the interaction is manifested by the appearance of a boundary
term in the string action. If the electromagnetic field is constant the new
term is gaussian and the path integral can be done exactly. Among the many
interesting properties that have been discovered by studying this simple
model, the appearance of extra massless states at certain critical values of
the electric \cite{ffi} and magnetic \cite{fer-por} fields could be
interpreted as signs of the existence of more symmetrical string phases.

Although very tractable, the open string case is far from being the most
realistic string-in-background model. One would like then to consider a
constant electromagnetic background, e.g. a constant magnetic field, as a
probe to reveal the properties of more fundamental models as closed strings
or some supersymmetric version of them. Unfortunately the pattern of
simplicity that gives rise to the exactly solvability of an open string in a
constant magnetic field is not repeated here because there is no possible
boundary coupling between a charged closed string and a magnetic background.
The string charge is now distributed through the internal points and the
interaction term is not a gaussian.

It is amazing that in spite of the above difficulty, an exactly solvable
model, which is a closed string analog of the open string in a constant
magnetic field, has been found \cite{russ-tse}. It turns out that a
conformal invariant model can be constructed if a Kaluza-Klein coordinate is
introduced, and the space curving effect due to the magnetic field is taken
into account.

Notwithstanding the low prospect of the open strings in backgrounds as
realistic models, they have been effective as the simplest systems where the
basic features of strings interacting with background fields can be studied.
Although, in particular, the open string propagating in constant
electromagnetic background has been extensively investigated in the
literature \cite{Fradkin}-\cite{Argyres}, not all its properties have been
properly understood. Hence, a thorough understanding of this case is still
needed to get insight on how to deal with the same problems in more complex
models with a similar background. This goal is the main motivation of the
present work.

In this paper we consider an open charged string propagating in a constant
magnetic background. We find the set of Casimir operators associated with
the algebra of the global symmetries, showing that two different linear
combinations of these invariants give rise to the string mass and rest
energy operators. The decomposition of the rest energy in terms of
invariants, one of which is the invariant string mass, provide us with a
systematic tool to identify and physically interpret each energy term in a
general way. This method can be easily extended to other string models. It
turns out that, at any background field strength, the string energy has the
same form as the Schwinger energy of a charged particle interacting with an
external magnetic field: $E_n^2=(2n+1)eH-geH\cdot S+m^2$. In addition to
this, the energy operator decomposition in terms of the physical invariants
provides us a consistent criterion for normal-ordering the energy. It also
shows that the normal order constant, used up to now in the literature \cite
{callan},\cite{nest},\cite{fer-por}, leads to an inconsistency. To determine
the correct normal ordering constant has important physical implications.
For instance, the normal ordering constant can in principle affect the value
of the Hagedorn temperature of a thermal string system,\cite{attick}, \cite
{russ}. Finally, with the aid of the energy invariant decomposition, we find
the operator of the spin projection along the magnetic field without taking
any approximation in the field strength. Using it and the correct string
mass we obtain one of the main results of this article: the Regge
trajectories of the system. As we will show below, the spin-magnetic field
interaction leads to the arising of a family of infinitely many new Regge
trajectories, each one associated with a different spin projection.

As has been shown in \cite{fe-in1}, a constant electromagnetic background
reduces the group of global symmetries of the open string to a subgroup of
Poincare. The associated Lie algebra of the conserved linear and angular
momenta, $\Pi _\mu $ and $M_{2n}\,_{2n+1},$ is given by the following
Poisson brackets

\begin{eqnarray}
\left\{ \Pi _\mu ,\Pi _\nu \right\} &=&qF_{\mu \nu ,}\qquad  \nonumber
\label{1} \\
\left\{ \Pi _\mu ,M_{2n}\,_{2n+1}\right\} &=&\eta _{\mu \;2n+1}\Pi
_{2n}-\eta _{\mu \;2n}\Pi _{2n+1,}  \label{1} \\
\left\{ M_{_{2n}\,_{2n+1}},M_{_{2m}\,_{2m+1}}\right\} &=&0.  \nonumber
\label{1}
\end{eqnarray}

\[
n=0,1,...,\frac{D-2}2
\]

The brackets (1) represent a subalgebra of Poincare with a central charge,
the central charge being given by the product of the total string charge $%
q=q_1+q_2$ and the background field strength $F_{\mu \nu }$.

Note that due to the constant electromagnetic background, the Lorentz
symmetry in (1) has been reduced to the subset of Lorentz transformations
which do not change the background field. One has no reason to expect that
in this case the Casimir operators are necessarily Lorentz scalars. Indeed,
using eqs.(1), it can be straightforwardly shown that the following set of
conserved quantities

\begin{equation}
I_0=\left( \Pi _0\right) ^2-\left( \Pi _1\right) ^2+2qM_{01}F^{01}  \label{3}
\end{equation}

\begin{equation}
I_i=-\left( \Pi _{2i}\right) ^2-\left( \Pi _{2i+1}\right)
^2+2qM_{2i\;2i+1}F^{2i\;2i+1}  \label{4}
\end{equation}

\[
i=1,...,\frac{D-2}2
\]
have zero brackets with the generators $\Pi _\mu $ and $M_{2n}\,_{2n+1}$,
and therefore they are the Casimir operators of the global symmetries (1).

Hereafter we consider a pure magnetic background, i.e., $F^{01}=0.$ Using
the operator formalism to quantize the theory in the standard light-cone
gauge, the quantum operators corresponding to the invariants (\ref{3}) and (%
\ref{4}) take the form

\begin{equation}
I_0=\frac 1{\alpha ^{\prime }}\sum\limits_{i=1}^{\frac{D-2}2}\left[
\sum\limits_{n=1}^\infty n\left( \overline{a}_n^ia_n^i+\overline{b}%
_n^ib_n^i\right) +\epsilon ^i\sum\limits_{n=1}^\infty \left( \overline{a}%
_n^ia_n^i-\overline{b}_n^ib_n^i\right) +\left| \epsilon ^i\right| \overline{a%
}_0^ia_0^i\right]  \label{nuev1}
\end{equation}

\begin{equation}
I_i=\frac{h_1^i+h_2^i}{\alpha ^{\prime }\pi }\left[ \sum\limits_{n=1}^\infty
\left( \overline{b}_n^ib_n^i-\overline{a}_n^ia_n^i\right) -\frac{\left|
\epsilon ^i\right| }{\epsilon ^i}\overline{a}_0^ia_0^i\right] ,  \label{12}
\end{equation}
where the following functions of the magnetic fields $H_i$,
\begin{equation}
\gamma ^i=\epsilon ^i-\frac{h_1^i+h_2^i}\pi ,  \label{11}
\end{equation}

\begin{equation}
\epsilon ^i=\frac 1\pi \left( \arctan h_1^i+\arctan h_2^i\right) ,  \label{7}
\end{equation}

\begin{equation}
h_a^i=\frac{q_a}TH_i,  \label{8}
\end{equation}
were introduced, and $\overline{a}_n^i(a_n^i),\overline{b}_n^i(b_n^i)$
denote quantum creation (annihilation) operators of the oscillator modes in
presence of the magnetic background (see ref.\cite{fe-in2}). They obey usual
commutation relations.

{}From (\ref{nuev1}), (\ref{12}), and using the results of refs.\cite{fe-in2},
\cite{nest}, it can be shown that the squared energy in the rest frame ($\Pi
_1=0$), as well as the string squared mass can be written as two independent
linear combinations of the invariants,

\begin{eqnarray}
\alpha ^{\prime }E^2 &=&\alpha ^{\prime }{\cal M}^2-\alpha ^{\prime
}\sum\limits_{i=1}^{\frac{D-2}2}I_i=\sum\limits_{i=1}^{\frac{D-2}2}\left[
\sum\limits_{n=1}^\infty n\left( \overline{a}_n^ia_n^i+\overline{b}%
_n^ib_n^i\right) \right.  \nonumber \\
&&\left. +\epsilon ^i\sum\limits_{n=1}^\infty \left( \overline{a}_n^ia_n^i-%
\overline{b}_n^ib_n^i\right) +\left| \epsilon ^i\right| \overline{a}%
_0^ia_0^i\right]  \label{13}
\end{eqnarray}

\begin{eqnarray}
\alpha ^{\prime }{\cal M}^2 &=&\alpha ^{\prime }\left( \sum\limits_{i=1}^{%
\frac{D-2}2}I_i+I_0\right) =\sum\limits_{i=1}^{\frac{D-2}2}\left[
\sum\limits_{n=1}^\infty n\left( \overline{a}_n^ia_n^i+\overline{b}%
_n^ib_n^i\right) \right.  \nonumber  \label{10} \\
&&\left. +\gamma ^i\sum\limits_{n=1}^\infty \left( \overline{a}_n^ia_n^i-%
\overline{b}_n^ib_n^i\right) +\frac{\left| \epsilon ^i\right| }{\epsilon ^i}%
\gamma ^i\overline{a}_0^ia_0^i\right]  \label{10}
\end{eqnarray}

Note, from eq.(\ref{13}), that in the presence of a magnetic background the
string rest energy is not simply equal to the string mass. As it will become
clear later, this is just the physical fact, already observed in the
particle case, that the energy contains contributions from orbital and spin
momenta interacting with a magnetic field. The rest energy decomposition in
terms of the invariant mass plus the other invariants $I_i$ will be basic
for this interpretation.

Operators (\ref{12}), (\ref{13}) and (\ref{10}) have order ambiguities due
to zero-point fluctuations and need to be normal ordered. Using the Riemann
zeta function

\begin{equation}
\zeta (z)=\sum\limits_{n=o}^\infty n^{-z}  \label{14}
\end{equation}

\begin{equation}
\zeta (-1)=-\frac 1{12}  \label{15}
\end{equation}
to regularize the divergent sums appearing due to normal order, one obtains
the following normal ordering constants

\begin{equation}
{\it c}_i=-\frac{h_1^i+h_2^i}{\alpha ^{\prime }2\pi }\frac{\left| \epsilon
^i\right| }{\epsilon ^i}\qquad ,  \label{nuev2}
\end{equation}

\begin{equation}
\alpha ^{\prime }{\it e}_0=\frac 12\sum\limits_{i=1}^{\frac{D-2}2}\left|
\epsilon ^i\right| -\frac{D-2}{24}\qquad ,  \label{nuev3}
\end{equation}

\begin{equation}
\alpha ^{\prime }{\it m}_0=\frac 12\sum\limits_{i=1}^{\frac{D-2}2}\frac{%
\left| \epsilon ^i\right| }{\epsilon ^i}\gamma _i-\frac{D-2}{24}\qquad ,
\label{nuev4}
\end{equation}
which have to be respectively added to eqs.(\ref{12}),(\ref{13}), and (\ref
{10}).

It should be observed, however, that the value of the zero-point fluctuation
contribution to the squared energy, eq.(\ref{nuev3}), differs from the one
commonly used for this system (see for instance refs.\cite{callan},\cite
{nest},\cite{fer-por}).The normal ordering constant of the squared energy up
to now accepted is

\begin{equation}
\frac 12\sum\limits_{i=1}^{\frac{D-2}2}\left[ \left| \epsilon ^i\right|
-\left( \epsilon ^i\right) ^2\right] -\frac{D-2}{24}  \label{21}
\end{equation}

Its derivation \cite{callan} was based in the argument that the zero-point
fluctuations should be defined in such a way to cancel an extra linear in n
c-number piece $\delta _{n+m,0}n\sum\limits_{i=1}^{\frac{D-2}2}\left[ \left|
\epsilon ^i\right| -\left( \epsilon ^i\right) ^2\right] ,$ which appears in
the Virasoro commutators $\left[ L_n,L_m\right] $. Such a prescription is
equivalent \cite{callan} to shift the normal ordering constant of the
Virasoro operator $L_{0\;}$from 1 to $1-\frac{\epsilon ^i}2\left( \frac{%
\left| \epsilon ^i\right| }{\epsilon ^i}-\epsilon ^i\right) $. Formally, the
zero-point fluctuation contribution (\ref{21}) can be found if, instead of (%
\ref{15}), the Riemann zeta-function

\begin{equation}
\zeta (z,q)=\sum\limits_{n=o}^\infty \left( n+q\right) ^{-z}  \label{23}
\end{equation}

\begin{equation}
\zeta (-1,q)=\frac q2(1-q)-\frac 1{12}  \label{24}
\end{equation}
is used to regularize the squared energy operator's divergent sums appearing
in the r.h.s. of (\ref{13}).

However, there is nothing wrong in keeping an extra linear in n c-number
term in the Virasoro commutators, since the usual Virasoro algebra already
has a linear in n c-number term. On the other hand, if one insists to use
the function (\ref{24}) to regularize the divergent sums, it is simple to
check that it leads to an inconsistency. When one applies (\ref{24}) to each
term of the combination $\alpha ^{\prime }{\cal M}^2-\alpha ^{\prime
}\sum\limits_{i=1}^{\frac{D-2}2}I_i$, and adds them to find the squared
energy ordering constant, the result does not coincide with the outcome of
the direct normal ordering of the r.h.s. of (\ref{13}), that is , with (\ref
{21}). Nevertheless, if the regularization is done through eq.(\ref{15}),
one obtains a compatible result, given in both cases by eq.(\ref{nuev3}). In
this way, the decomposition of the squared energy in terms of the physical
invariants ${\cal M}^2$ and $I_i$ provides a consistent criterion to find
the correct normal ordering constant. We expect that a similar approach will
give the appropriate ordering constant in the case of the closed string in a
magnetic background.

Let us find the physical meaning of $\sum\limits_{i=1}^{\frac{D-2}2}I_i$.
With this aim, consider $\epsilon ^i\geq 0$ ( a similar analysis can be done
if $\epsilon ^i<0$ ). Using the operators (\ref{12}),(\ref{10}) and (\ref{13}%
) in their normal order form, and writing the normalized magnetic field
components $h_a^i$ in terms of $H^i$

\begin{equation}
\frac{h_1^i+h_2^i}{2\pi }=\frac{q_1H^i}{2\pi T}+\frac{q_2H^i}{2\pi T}=\alpha
^{\prime }qH^i  \label{26}
\end{equation}
we obtain for the squared rest energy

\begin{equation}
\alpha ^{\prime }:E^2:=\alpha ^{\prime }\sum\limits_{i=1}^{\frac{D-2}%
2}\left[ (2\overline{a}_0^ia_0^i+1)qH^i-2qH^i\sum\limits_{n=1}^\infty \left(
\overline{b}_n^ib_n^i-\overline{a}_n^ia_n^i\right) \right] +\alpha ^{\prime
}:{\cal M}^2:  \label{27}
\end{equation}

Comparing eq.(\ref{27}) with the Schwinger energy spectrum of a charged
particle of spin $S$, charge $e$, and mass $m$

\begin{equation}
E_n^2=(2n+1)eH-ge{\bf H}\cdot {\bf S}+m^2  \label{25}
\end{equation}
we see that for each string state the operator $(2\overline{a}%
_0^ia_0^i+1)qH^i$ corresponds to the Landau level term of the particle,
while the term $2qH^i\sum\limits_{n=1}^\infty \left( \overline{b}_n^ib_n^i-%
\overline{a}_n^ia_n^i\right) $ is the string version of the spin-field
interaction contribution with a tree level gyromagnetic ratio $g=2$ ( in
agreement with ref.\cite{fer-por-tel}) for all string states. Hence, the
operator of the spin projection along the magnetic field is

\begin{equation}
S_H^i=\sum\limits_{n=1}^\infty \left( \overline{b}_n^ib_n^i-\overline{a}%
_n^ia_n^i\right)  \label{28}
\end{equation}
and we can conclude that the invariants' sum $\sum\limits_{i=1}^{\frac{D-2}%
2}I_i$ contains the contributions of Landau levels and spin-field
interaction, that is, orbital- and spin-background field interactions, to
the squared energy.

Some remarks are in order. The rest energy decomposition in function of the
invariants $I_i$ and the physical string mass ${\cal M}^2$ which is an
invariant too, has made it possible to establish a term by term
correspondence between the squared energies of string and particle. Each
string energy term has a magnetic field dependent coefficient which exactly
coincides with the coefficient of the associated particle term at any given
background field strength. If the energy is not written in terms of the
correct invariant string mass, that is, as ${\cal M}^2-\sum\limits_{i=1}^{%
\frac{D-2}2}I_i$, it becomes necessary to assume \cite{fer-por} a weak field
limit approximation, $\alpha $'$eH\ll 1$, in order to obtain a term by term
identification and equal coefficients for string and particle energies. In
this particular limit one has $\gamma ^i\simeq 0$, and of course the
invariant mass simply reduces to the free string mass.

Let us discuss some peculiarities of the energy spectrum of the theory. For
the sake of simplicity we may assume that the magnetic field is nonzero only
along a given i-direction. In this case, only one of the $\frac{D-2}2$
magnetic components $H^i$ is present. Physical states are generated by
acting with the creation operators on the ground state, which we will
consider, following \cite{callan}, an eigenstate of the zero mode associated
with the center of mass. The spin-one state $\overline{a}_1\left|
0\right\rangle $, with spin projection antiparallel to the field, has mass
eigenvalue

\begin{equation}
\alpha ^{\prime }:{\cal M}^2:\overline{a}_1\left| 0\right\rangle =\alpha
^{\prime }{\sl m}^2\overline{a}_1\left| 0\right\rangle =\frac 32\gamma
\overline{a}_1\left| 0\right\rangle \; ,  \label{31}
\end{equation}
where we have suppressed the i-label, and rest energy

\begin{equation}
\alpha ^{\prime }:E^2:\overline{a}_1\left| 0\right\rangle =\alpha ^{\prime }%
{\sl \varepsilon }^2\overline{a}_1\left| 0\right\rangle =\frac 32\epsilon
\overline{a}_1\left| 0\right\rangle  \label{32}
\end{equation}

Since for $\epsilon >0,$ one has $\gamma <0$ for any non zero magnetic
field, this state has negative squared mass but positive squared energy. We
call it a stable tachyon.

The opposite situation, i.e. states with positive squared mass and negative
squared energy is also present. These are the type of instabilities found in
\cite{fer-por}. They also appear in the context of the standard model, \cite
{niel-ole}, \cite{ambj},where the gyromagnetic ratio is $g_W=2$.

Another important property of the energy spectrum of the charged string in a
magnetic background is the modification of the Regge trajectories. Once we
identify the string mass and the spin projection, it is simply to see that
zero Landau level states, lying in the first Regge trajectory satisfy

\begin{equation}
S=\alpha ^{\prime }{\sl m}^2+1-\frac \gamma 2+\gamma S_H  \label{33}
\end{equation}
where $S$ is the spin of the state, ${\sl m}$ is the mass and $S_H$ is the
spin projection along the magnetic field.

Note that the first Regge-trajectory has been transformed in a family of
infinitely many lines, one for each value of the spin projection. The
magnetic field has broken the degeneracy of these states.

If we compare eq.(\ref{33}) with the Regge-trajectory of the free-string, it
is evident that for states with spin projection $S_H=0$, the effect of the
magnetic background reduces just to an increase in the $S-$intercept from 1
to $1-\frac \gamma 2$. A direct consequence of this shift is that the
free-string massless state acquires a mass in the presence of a magnetic
background, otherwise it would have an arbitrary field-dependent spin $%
1-\frac \gamma 2$.

As we have previously observed, states with a given $S_H$ belong to one line
of the family. Therefore, states with different projections of a given spin
have different masses. It is clear from eq.(\ref{33}) that as the magnetic
field increases, the squared masses of states with negative $S_H$ decrease,
while the squared masses of states with positive $S_H$ increase.

In the rest frame, the energy eigenvalues for zero Landau level states
belonging to the first Regge-trajectory satisfy

\begin{equation}
S=\alpha ^{\prime }{\sl \varepsilon }^2+1-\frac \epsilon 2+\epsilon S_H
\label{34}
\end{equation}

As the magnetic field increases, the squared energy decreases for states
with positive $S_H$ and increases for states with negative $S_H$.

Therefore, we have shown that the Regge trajectories of a charged string are
non-trivially modified in a constant magnetic background. The presence of a $%
S_H$-dependent term splits each one of the Regge trajectories, which
characterize the free string model, in a family of lines labelled by the
spin projection eigenvalue.

In this paper the global symmetries of the charged string in a constant
magnetic background have been used as a tool to deepen in the physical
understanding of the system. All the results found here are the outcome of
expressing the energy as a combination of the invariant string mass and the
invariant operators $I_i$. A similar strategy can be followed to deal with
closed \cite{russ-tse} and heterotic strings \cite{russ-tse2} in a constant
magnetic background. In particular, a proper mass identification within the
energy expression can be crucial in relation with recent claims, \cite
{russ-tse2}, \cite{russ}, that, at certain critical value of the magnetic
field, infinitely many extra massless states arise, a sign that might
indicate a restoration of spontaneously broken symmetries.

{\bf Acknowledgment }

One of the authors, V.I., would like to thank A.A. Tseytlin for useful
discussions.

This work has been supported in part by a National Science Foundation Grant
under Contract No. PHY-9414509.

\end{document}